\documentclass[aps,prd,showpacs,nofootinbib,notitlepage, twocolumn]{revtex4-1}
\usepackage{latexsym}
\usepackage{amsbsy,amsmath,amsthm,amsfonts,calc,amssymb,graphicx,accents,mathrsfs}
\usepackage{braket,textcomp,upgreek}
\usepackage[percent]{overpic}
\usepackage{psfrag}
\usepackage{graphicx}
\usepackage{caption,subcaption}
\usepackage{color}
\usepackage{psfrag}
\usepackage{enumerate}
\theoremstyle{plain}

\theoremstyle{definition}
\DeclareFontFamily{U}{rsfs}{}         
\DeclareFontShape{U}{rsfs}{m}{n}{<5> rsfs5 <6><7> rsfs7          %
	<8><9><10><10.95><12><14.4><17.28><20.74><24.88> rsfs10}{}     %
\DeclareMathAlphabet{\mathfs}{U}{rsfs}{m}{n}                     %
\definecolor{indiagreen}{rgb}{0.07, 0.53, 0.03}
\def\beq{\begin{eqnarray}}
\def\eeq{\end{eqnarray}}

\def\nn{\nonumber\\}

\def\={\stackrel{\Delta}{=}}

\def\lie{\pounds}

\begin{document}
	\begin{flushleft}
		\footnotesize USTC-ICTS/PCFT-21-24
	\end{flushleft}
	\title{Inducing supertranslations: The membrane picture}
	\author{Avirup Ghosh}\email{avirup@ustc.edu.cn}
	\affiliation{Interdisciplinary Center for Theoretical Study, University of Science and Technology of China, and 
		Peng Huanwu Center for Fundamental Theory, Hefei, Anhui 230026, China	}
	\begin{abstract}
	Dynamical evolution of black holes may be studied perturbatively in the membrane picture. We study the case of a body freely falling into a black hole, in the Rindler approximation. The motivation is to see if this process is accompanied by the induction of a non-zero supertranslation. Following previous studies, a notion of induction of supertranslation during a dynamical evolution is first introduced and then, the case of a freely falling body is investigated. 
	\end{abstract}
	\maketitle
\section{Introduction}
The first demonstration of the enhancement of symmetries at the boundary of a spacetime, which we now refer to as asymptotic symmetries, was in the case of asymptotically flat spacetimes at null infinity \cite{Bondi:1962px,Sachs:1962wk,Geroch:1977jn}, leading to the Bondi-Metzner-Sachs ($BMS$) group. Later this was adopted for asymptotically $AdS$ spacetime \cite{Brown:1986nw} and remains a corner stone in the understanding of the AdS-CFT correspondence.

In recent times there has been much interest in the symmetries near the horizon of a stationary black hole, primarily owing to a suggestion that these account for extra hair that black holes carry which could resolve the information loss paradox \cite{Hawking:2016msc}. The original idea about studying near horizon symmetries however dates back to  \cite{Carlip:1999cy}. The main aim then was to reproduce the Bekenstein -Hawking entropy formula \cite{Bekenstein:1973ur,Hawking:1971vc} by counting the states in a representation of an effective conformal symmetry on the horizon 	\cite{Hotta:2000gx,Koga:2001vq}. In recent times, however, the general interest has been to find $BMS$ like symmetries near the horizon \cite{Donnay:2016ejv,Donnay:2015abr,Averin:2016ybl,Afshar:2016wfy,Setare:2016jba,Mao:2016pwq,Ciambelli:2019lap,Carlip:2017xne,Sousa:2017auc,Grumiller:2019fmp,Adami:2020amw,Grumiller:2020vvv} or null surfaces in general \cite{Chandrasekaran:2018aop,Adami:2020ugu}. The main aim of this paper wont be along these lines but an attempt to study dynamical black hole spacetime in relation to these symmetries. By this we however do not mean that we would like to find out $BMS$ like symmetries near a dynamically evolving horizon, for there may not be any. To understand what we would like to explore let us recall the symmetry group at null inifinty. 

In the case of asymptotically flat spacetimes at null infinity the usual group of isometries, the Poincare group gets enhanced to the $BMS$ group which is a semidirect product of the infinite dimensional abelian subgroup of supertranslations and the Lorentz group \footnote{There could be a further enhancement of the Lorentz group as well depending on the boundary conditions \cite{Barnich:2009se,Campiglia:2014yka,Campiglia:2015yka}}. We will be interested in this abelian subgroup of supertranslations only. Their action on Minkowski spacetime  corresponds to angle dependent translations and thus normal translations form a subgroup of the supertranslations. The conserved charges corresponding to the generators, on the phase space of gravitational degrees of freedom,  however vanish (except for finitely many). This raises doubts whether they are indeed physical or continue to be gauge. Their relevance however transpires when one studies dynamical processes in the bulk.

Suppose we have a dynamical process in the bulk such that a gravitational wave is emitted and reaches the future null infinity. Then the epochs, prior to and after the passage, of the gravitational wave are stationary. In both these epochs there is a natural choice of frames adapted to the stationarity. It can then be shown that these two frames are related to each other by a supertranslation. This is precisely what tells us that the supertranslations are physical \cite{Strominger:2014pwa,Hollands:2016oma}.

At the horizon a similar enhancement occurs. But in this case one obtains the semidirect product of the supertranslations and the group of isometries of the horizon cross section \footnote{Here also there could be a further enhancement depending on the boundary conditions \cite{Chandrasekaran:2018aop}.}. As is the case with supertranslations at the null infinity the charges are trivial, except for finitely many. Therefore one contemplates whether a similar dynamical calculation can be done in this case as well. The challenges here are plenty. Unlike in the asymptotically flat case this is a strong gravity regime and calculating the fields, of isolated sources, near the horizon can be tricky. In particular a perturbative treatment may not be feasible. A non perturbative approach on the other hand requires the full utilisation of the numerical relativity machinery. To understand if an analytic calulation can  be done let us briefly mention the approach to be taken. 

One first has to identify a scalar quantity, defined on a stationary horizon, that transforms under a supertranslation. In our case this will be taken to be the divergence of the rotation one form. By using the supertranslation freedom one may then set this quantity to be zero for a stationary horizon. The evolution of this quantity along the horizon must then be taken, when the black hole is dynamically evolving. If there is a change in this quantity then we must say that a supertranslation has been induced else not.

To initiate this we first need a description of dynamically evolving horizons. There are essentially two such descriptions. The first being a `\emph{Dynamical horizon}(DH) which is a space-like three surface foliated by marginally trapped surfaces \cite{Ashtekar:2003hk}. This description is inherently non perturbative. Hence in general a full simulation of the Einstein's equation is required to study the evolution of the divergence of the rotation one form along a DH. However very special cases such as a spherically symmetric black hole evolving into a spherically symmetric black hole can be studied analytically as has been done in \cite{Chatterjee:2020enf}.

Another approach is the null event horizon (EH), which in the dynamical case is expanding and shearing as opposed to being non expanding in the stationary case. In this case a perturbative treatment may be taken and the expansion is in powers of parameters of the perturbing or infalling object. This forms the main content of the membrane paradigm framework. In the membrane picture the evolution of quantities along a time-like membrane just outside the horizon is considered. The membrane replaces the horizon and mimicks a black hole horizon. The event horizon is just the null limit of this membrane \cite{Thorne:1986iy,Price:1986yy,Suen:1988kq}. In this paper we will be interested in this second approach as opposed to the Dynamical horizon approach. This is because the symmetry assumptions in the previous case simplifies the evolution equation significantly and the effect of certain terms in the evolution is completely missed. For example by assuming spherical symmetry the shear in the direction tranverse to the DH is completely eliminated and the only term responsible for inducing the supertranslation is a stress energy term appearing in the evolution equation. Hence one may interpret it as a purely matter induced supertranslation. We would however like to explore supertranslations which are not necessarily matter induced and thus need to retain terms lost due to symmetry assumptions.

\section{Inducing supertranslations}
\subsection{Action of super-translations on a stationary event horizon}
In this section we will see how certain geometric quantities, associated with a stationary event horizon, transform under the action of a supertranslation. The coordinates on the cross section of the horizon are denoted by $\tau^A$, the parameter along the null generators ($l$) is denoted by $v$ and the parameter along the other null vector ($n$) is denoted by $r$. The rotation one form $\omega_A:=-g(n,\nabla_{\partial_A}l)$ and the transverse extrinsic curvature $K^{(n)}_{AB}:=g(n,\nabla_{\partial_A}\partial_B)$ are the two geometric quantities, relevant to a stationary EH, that transform under a supertranslation \cite{Ghosh:2020wjx}. Let us denote the stationary event horion by $\mathcal H$. A supertranslations is then a map $\psi:\mathcal H \rightarrow\tilde{\mathcal H}$, that can be represented as $\tilde v=v+\mathcal F, ~\tilde\tau^A=\tau^A, ~\tilde r=r$. In terms of the connection we have,
\begin{gather}
\kappa=-g(n,\nabla_ll)=\Gamma^{\mathfrak a}_{vv}g_{\mathfrak a r}~~~\omega_A=\Gamma_{Av}^{\mathfrak a}g_{\mathfrak a r},
\end{gather}
where we have denoted the surface gravity by $\kappa$ and the $4$ dimensional spacetime coordinates as $\mathfrak a$ etc. On Using the transformation rules of a Levi-Civita connection \footnote{$
	\tilde \Gamma^{\tilde{\mathfrak a}}_{\tilde{\mathfrak b}\tilde{\mathfrak c}}=\frac{\partial \tilde X^{\tilde{\mathfrak a}}}{\partial X^{\mathfrak a}}\frac{\partial X^{\mathfrak b}}{\partial \tilde X^{\tilde{\mathfrak b}}}\frac{\partial X^{\mathfrak c}}{\partial \tilde X^{\tilde{\mathfrak c}}}\Gamma^{\mathfrak a}_{\mathfrak b \mathfrak c}+\frac{\partial^2 X^{\mathfrak a}}{\partial \tilde X^{\tilde{\mathfrak b}}\partial \tilde X^{\tilde{\mathfrak c}}}\frac{\partial \tilde X^{\tilde{\mathfrak a}}}{\partial X^{\mathfrak a}}$}
we have,
\begin{gather}
\tilde\kappa=\tilde\Gamma^{\tilde{\mathfrak a}}_{\tilde v\tilde v}\tilde g_{\tilde{\mathfrak a} \tilde r}=\Gamma^{\mathfrak a}_{vv}g_{\mathfrak a r}=\kappa\nn
\tilde\omega_{A}=\tilde\Gamma^{\tilde{\mathfrak a}}_{A\tilde v}\tilde g_{\tilde{\mathfrak a} \tilde r}=\Gamma^{\mathfrak a}_{Av}g_{a r}+\Gamma^{a}_{vv}g_{a r}~\partial_A\mathcal F=\omega_A+\kappa~\partial_A\mathcal F\\
\tilde K^{(n)}_{AB}=K^{(n)}_{AB}-\bigg[\mathcal D_{(A}\mathcal D_{B)}\mathcal F+\mathcal D_{(A}\mathcal F~\omega_{B)}+\omega_{(A}\mathcal D_{B)}\mathcal F\nn
+\mathcal \kappa~\mathcal D_{(A}\mathcal F~\mathcal D_{B)}\mathcal F\bigg]\label{ECT},
\end{gather}
where $\mathcal D$ denotes the covariant derivative on the cross-sections of $\mathcal H$.
%
%
%
%
The supertranslation freedom, being a symmetry, can be used to render the rotation one form divergence free. To  see it, note that under a supertranslation the metric on the cross-section, thus the covariant derivative compatible with it remain invariant. Thus under a supertranslation,
\begin{gather}
\mathcal D^A\tilde{\omega}_A=\mathcal D^A \omega_A+\kappa \mathcal D^2 \mathcal F
\end{gather}
By an appropriate choice of $\mathcal F$ one therefore can render $\omega_A$ divergence free. This is equivalent to choosing a preferred foliation of the horizon. The condition and therefore the choice of foliation is preserved along $\mathcal H$.
During the course of a dynamical evolution this however may not be true. We may therefore say that during a dynamical process a supertranslation has been induced if the divergence free condition is preserved, else not. This approach is closely related to the notion advocated in \cite{Rahman:2019bmk}.
\subsection{Evolution equation}
In the event horizon or the membrane paradigm framework the dynamics of a black hole is captured by the evolution of the null event horizon which unlike the stationary event horizon is both expanding and shearing. As the spacetime is perturbed due to ingoing flux the event horizon starts expanding, from the initial bifurcaion surface, even before the flux has fallen across it. The expansion decreases during and after the passage of matter across the horizon, finally becoming zero at asymptotic future. This indicates that a stationary state has been reached. We will try to evaluate the total change in the divergence of the rotation one form along the event horizon.
\begin{gather}
\nabla_l\omega^A=q^{AB}\bigg[\mathcal D_C\big(q^{CD}K^{(l)}_{DB}\big)+R(\tau,\partial_B)- \mathcal D_{\partial_B}K^{(l)}\nn
+\omega_B~K^{(l)}+\partial_B\kappa_l
+2\omega^C K^{(l)}_{BC}\bigg]\label{omegaev}
\end{gather}

The evolution of the divergence is captured by the following equation,
\begin{gather}
\lie_X(div~ \omega)=\mathcal D_A(\nabla_l~\omega^A)+\omega^A\partial_AK^{(l)}
\end{gather}
The above equation contains terms involving the expansion and shear of the null generators. Thus we need expressions for their evolution as well.
Decomposing the extrinsic curvature into a trace and a trace free part, as $K^{(l)}_{AB}:=\frac{K^{(l)}}{2}q_{AB}+\Sigma^{(l)}_{AB}$ we have the following equations for their evolution.
\begin{gather}
\nabla_lK^{(l)}-\kappa K^{(l)}=K^{(l)}_{AB}h^{AC}h^{BD}K^{(l)}_{CD}+R(l,l)\nn
\nabla_l\Sigma^{(l)}_{AB}-\kappa\Sigma^{(l)}_{AB}=g(l,C(l,\partial_A)\partial_B)-\frac{h_{AB}}{2}\Sigma^{(l)}_{CD}\Sigma^{(l)AB}\nn
-\Sigma^{(l)}_{AC}\Sigma^{(l)C}_{~~~B}
\end{gather}
Note that the trace and trace free parts defined here differ by a negative sign from the usual shear and expansion i.e $\Sigma_{AB}=-\sigma_{AB}$ and $K=-\theta$. The evolution of the induced metric thus carries an extra negative sign $\nabla_l~q_{AB}=-2K^{(l)}_{AB}$. Having obtained the evolution equations we will now consider a  particle freely falling into the horizon. To further simplify the calculation we will employ the Rindler approximation of a black hole.

\subsection{The source}
To understand the field of the source near the horizon we first need to recall the Rindler approximation for a black hole horizon. Let us take the example of the Schwarschild black hole,. It is well known that in the near horizon limit the Schwarzchild metric looks like,
\begin{gather}
ds^2=-r^2~d\tau^2+\frac{dr^2}{\kappa^2}+4M^2d\Omega^2
\end{gather}
By a redefinition of coordinates,
\begin{gather}
x = 2M\sin(\theta_0) \phi,~~y =2M (\theta -\theta_0),~~Z = \frac{r}{\kappa},\nn
t = Z\sinh(\kappa \tau),~~z = Z \cosh (\kappa \tau),
\label{trans}
\end{gather}
we can then write the metric as a Minkowski metric.  The horizon here is the surface $t=z$ and the approximation holds only within a small solid angle around $\theta=\theta_0,~\phi=0$. In the Minkowski coordinates the horizon cross section matches with the cross section of the Rindler horizon only within a region $x^2+y^2<a^2$ say, depending on the choice of $M$.

We will consider a freely falling particle located at $x=y=0,~z=z_0$ falling across the Rindler horizon. To set up the problem let us consider the metric for a particle of mass $m$ located at $x=y=0$ and $z=z_0$, in isotropic coordinates. The coordinate is comoving with the particle and thus the metric is time independent and equivalent to the Schwarzschild metric in isotropic coordinates.
\begin{gather}
ds^2 = -\frac{\left(1-\frac{ m }{ 2\sqrt{ \rho^2 + (z-z_0)^2 } }  \right)^2}{\left(1+\frac{ m }{ 2\sqrt{ \rho^2 + (z-z_0)^2 } }  \right)^2} dt^2 
\nn
+\left(1+\frac{ m }{ 2\sqrt{ \rho^2 + (z-z_0)^2 } }  \right)^4\left( dz^2 + dx^2 + dy^2 \right),
\end{gather}
where $\rho^2 = x^2+y^2$. We must calculate the non-zero components of the electric part of the Weyl tensor on the horizon.  Since the metric is flat, at zeroth order, the outgoing null vector $\accentset{(0)}{l}\equiv (-1,1,0,0)$ are the generators of the horizon. If $s$ is a parameter along the generator then in parametric form the equation for the surface is $t=s,~z=s,~X=x,~Y=y$ (where $s,X,Y$ are coordinates intrinsic to the horizon) such that $s=0$ is the bifurcation surface. However note that the generator here is affinely parametrised and therefore generates the flat space light cones and not the Rindler horizon. In order to find the generators of the Rindler horizon one simply has to boost this null vector. One can easily check that $X=f~\accentset{(0)}{l}$ with $f=\kappa s$, where $\kappa$ is a constant along the generators, generates the Rindler horizon, of accelaration parameter $\kappa$. If the parameter along $X$ is taken to be $v$ i.e $X=\partial_v$, then $s=Ce^{\kappa v}$ for some constant $C$ which has to be appropriately chosen. In this case if we assume that the particle crosses the horizon at $v=0$ where $z=z_0$, then $C=z_0$. It will however not be necessary to reparametrise the geodesic as we will be doing all our calculations in the affine parametrisation. Thus what we have here is a Rindler horizon in affine parametrisation, the bifurcation surface being located at $v\rightarrow -\infty$.

The generators as well as the embedding functions get modified at higher order. However while calculating the changes in horizon quantitites at order $m$ the unperturbed value of the generators and horizon parametrisation can be used as corrections to these only appear at order $m^2$. In the foregoing discussion we will separate the sections involving calculation linear in $m$ and quadratic in $m$.
\subsection{Linear order in $m$}\label{order1}
\begin{figure}[h]
	\includegraphics[width=8cm]{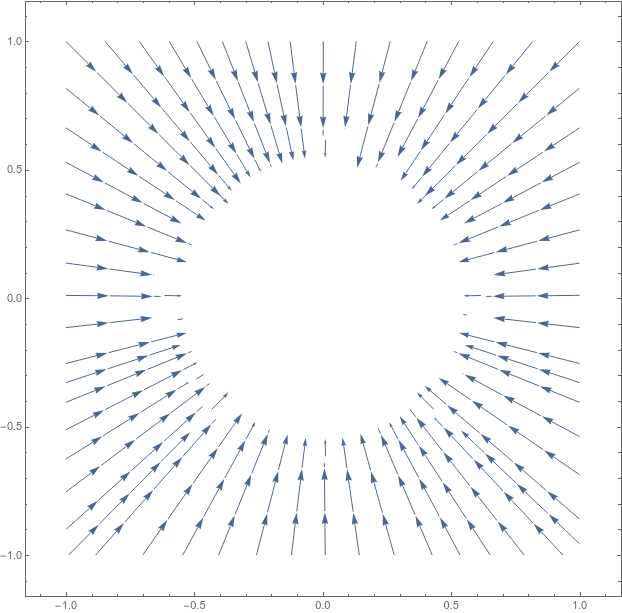}
	\caption{Plot of $\omega^A$ on the $x,y$ plane at the initial slice $s\rightarrow 0$}
	\label{omega}
\end{figure}

We now need an expression for the Tidal forces. Since the Weyl tensor is already of order $m$ or higher it is enough to take zeroth order values of $l$ and the embedding functions. Thus on the dynamical event horizon $\mathscr H$ (say) the values of components of the Weyl tensor are,
\begin{gather}\label{exactweyl}
g(\accentset{(0)}{l},\accentset{(1)}{C}(\accentset{(0)}{l},\partial_X)\partial_X) = - g(\accentset{(0)}{l},\accentset{(1)}{C}(\accentset{(0)}{l},\partial_Y)\partial_Y)\nn 
=\frac{-3 m \big(X^2-Y^2\big)}{\big(X^2+Y^2+(s-z_0)^2\big)^{\frac{5}{2}}},\nn
g(\accentset{(0)}{l},\accentset{(1)}{C}(\accentset{(0)}{l},\partial_Y)\partial_X)
=-\frac{6 m ~X Y}{\big(X^2+Y^2+(s-z_0)^2\big)^{5/2}}.
\end{gather}
To find the expansion and the shear on the horizon we use the Newman -Penrose equations to linear order in $m$,
\begin{gather}
\frac{d\accentset{(1)}{\Sigma}_{AB}}{ds}=g(\accentset{(0)}{l},\accentset{(1)}{C}(\accentset{(0)}{l},\partial_A)\partial_B).
\end{gather}
As has been discussed before, the boundary condition here should be at asymptotic future. This is due to the teleological nature of the event horizon whereby one must assume that the shear and the expansion goes to zero in the future when the black hole becomes stationary. This gives the following expressions for the shear $\Sigma_{AB}$
\begin{gather}
\accentset{(1)}{\Sigma}_{XX}=\frac{m (X^2-Y^2) \left(\frac{(s-z_0) \left(2 (s-z_0)^2+3 \left(X^2+Y^2\right)\right)}{\left((s-z_0)^2+X^2+Y^2\right)^{3/2}}+2\right)}{\left(X^2+Y^2\right)^2}\nn
\accentset{(1)}{\Sigma}_{XY}=\frac{2 m X Y \left(\frac{(s-z_0) \left(2 (s-z_0)^2+3 \left(X^2+Y^2\right)\right)}{\left((s-z_0)^2+X^2+Y^2\right)^{3/2}}+2\right)}{\left(X^2+Y^2\right)^2}\nn
\accentset{(1)}{\Sigma}_{YY}=\frac{m (Y^2-X^2) \left(\frac{(s-z_0) \left(2 (s-z_0)^2+3 \left(X^2+Y^2\right)\right)}{\left((s-z_0)^2+X^2+Y^2\right)^{3/2}}+2\right)}{\left(X^2+Y^2\right)^2}
\end{gather}
We will not take any approximations here as in \cite{Suen:1988kq} but will straightaway calculate the change in the divergence of the rotation one-form $\omega_A$. To obtain the change in the rotation one form let us consider the equation for it's evolution at linear order,
\begin{gather}
\frac{\partial \accentset{(1)}{\omega}^X}{\partial s}=\partial_X\accentset{(1)}{\Sigma}_{XX}+\partial_Y\accentset{(1)}{\Sigma}_{YX}\nn
\frac{\partial\accentset{(1)}{ \omega}^Y}{\partial s}=\partial_Y\accentset{(1)}{\sigma}_{YY}+\partial_X\accentset{(1)}{\Sigma}_{XY}
\end{gather}
Note that the trace part $K^{(l)}$ has been taken to be zero at this order. This is because we have assumed the stress energy tensor to be zero. The boundary conditions that $\omega^A$ is zero at the final slice then gives the following solutions,
\begin{gather}
\accentset{(1)}{\omega}^X=-\frac{m X}{\left(X^2+Y^2+(s-z_0)^2\right)^{3/2}}\nn
\accentset{(1)}{\omega}^Y=-\frac{m Y}{\left(X^2+Y^2+(s-z_0)^2\right)^{3/2}}
\end{gather}
A plot of these values at the initial slice is shown in fig. (\ref{omega}). Further,
\begin{gather}
div ~\accentset{(1)}{\omega}=\frac{3 m \left(X^2-Y^2\right)}{\left(X^2+Y^2+(s-z_0)^2\right)^{5/2}}
\label{divomegalimit}
\end{gather}
\begin{figure}[h]
	\includegraphics[width=8cm]{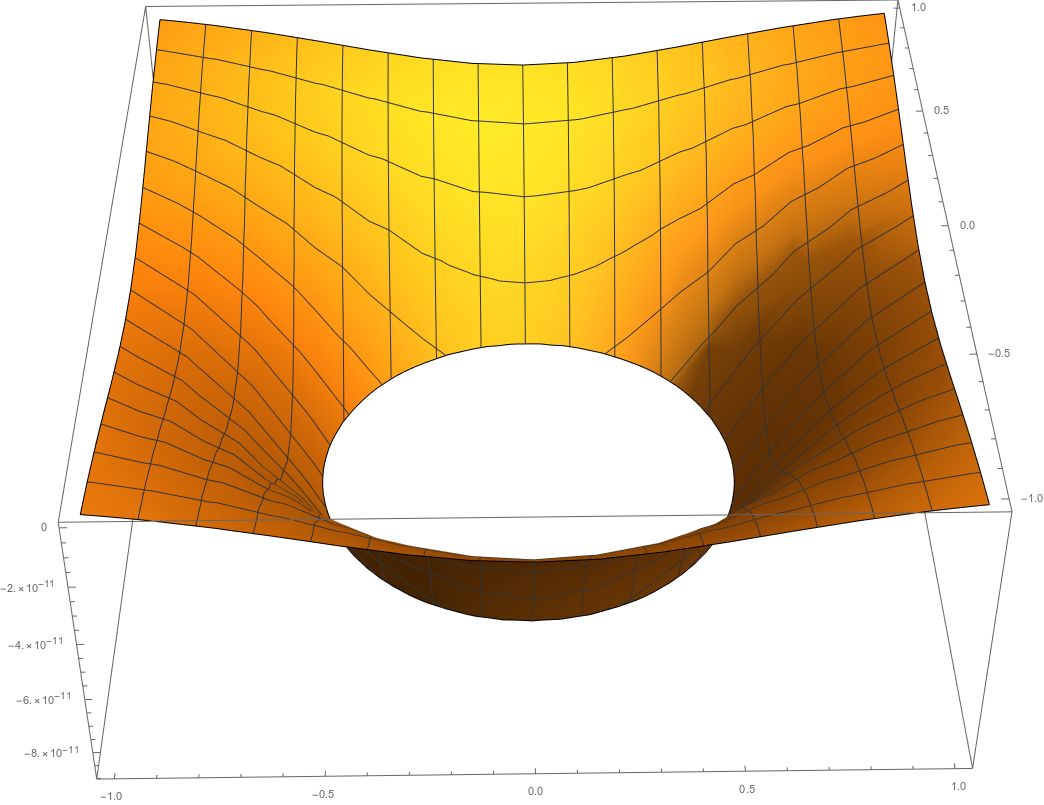}
	\includegraphics[width=8cm]{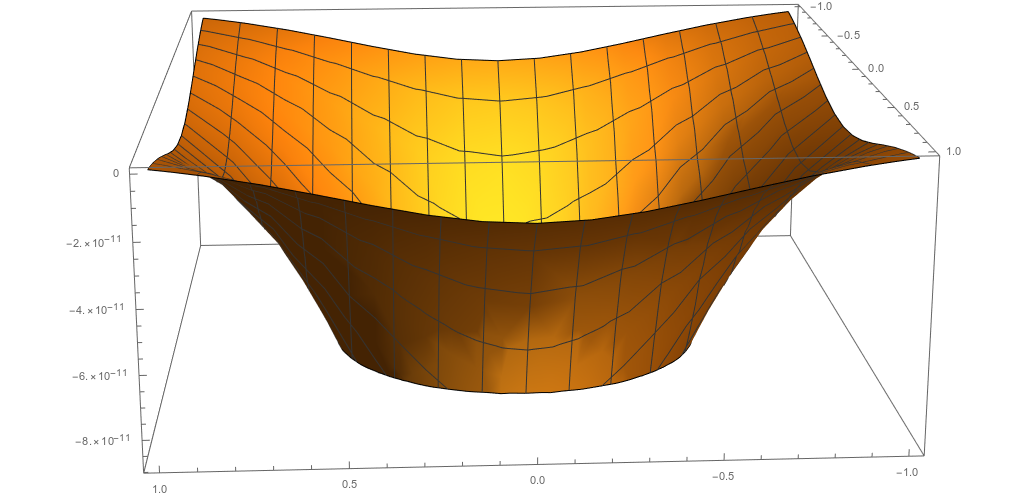}
	\caption{Plot of divergence of $\omega^A$ on the $x,y$ plane at the initial slice $s\rightarrow 0$}
	\label{divomega}
\end{figure}
The divergence at the initial slice has also been plotted in fig. (\ref{divomega}). Note that for both these plots we have excised a small region $X^2+Y^2<b^2$ where $b$ is small. The choice of parameters here are $b=0.3,~m=10^{-10},z_0=1$. The range of $x,y$ has been taken to be $(-1,1)$. The range here is flexible and can be modified by appropriately choosing $M$ eq. (\ref{trans}). However $M$ must satisfy the bound derived in \cite{Suen:1988kq,Amsel:2007mh,Bhattacharjee:2014eea} so that casutics do not form to the future of the bifurcation surface. If $b$ is roughly taken to be the radius of the body, then this removes the generators, that pass through the interior of the body, from the plot. Since we have assumed a zero stress energy tensor and the metric exterior to a particle as our source, the results obtained are true only for those generators that do not pass through the body.  This will be apparent at order $m^2$ where the results will be seen to diverge at $x=y=0$. We will only be interested in the geometric quantities of these generators which do not pass through the body itself. This is because we want to study the effects of non matter like terms in the evolution equation. 

Note that boundary conditions imposed on  $\omega^A$ and its divergence are are at the asymptotic future. The reason for this is not the teleological property of $\mathscr H$, but rather the boundary conditions to be imposed at order $m^2$. It is worth observing that $\omega_A$ can also be calculated by using expressions for $l$, the covariant derivative and $n$ corrected to order $m$. However to avoid hassle one uses the evolution equation. The linear order effects in $l$ etc. are correctly taken care of by the linear order terms in the Weyl tensor. In the next section we are going to find and use these linear order corrections. It is therefore important to ensure that the results obtained using the evolution equation are consistent with those obtained from latter considerations. The boundary conditions have been assumed foreseeing this requirement. 

A look at the figures clearly reveal that even in this very simple example there is a non zero change in the divergence of the rotation one-form implying that for generic processes a supertranslation will be induced.
\subsection{Quadratic order in $m$}\label{order2}
Let us recall the evolution equation for the divergence. At linear order only the term $\mathcal D^A\mathcal D^B\Sigma_{AB}$ contributed to this equation as every other term is of quadratic or higher order. The effect of $K^{(l)}$ e.g was therefore hidden at this order and appears only at order $m^2$. Apart from this a variety of terms enter the picture. Before discussing about each contributing term, let us first gather the ingredients. The expansion e.g can be obtained by solving the differential equation equation for $K^{(l)}$ with prescribed boundary conditions.
\begin{gather}
\accentset{(2)}{K}^{(l)}=\frac{2 m^2}{\left(X^2+Y^2\right)^2} \bigg[\frac{(s-z_0) \left(X^2+Y^2\right)^2}{4 \left(X^2+Y^2+(s-z_0)^2\right)^2}\nn
+\frac{15 (s-z_0) \left(X^2+Y^2\right)}{8 \left(X^2+Y^2+(s-z_0)^2\right)}
-\frac{4 \left(X^2+Y^2+2 (s-z_0)^2\right)}{\sqrt{X^2+Y^2+(s-z_0)^2}}\nn
-\frac{15}{8} \sqrt{X^2+Y^2} \tan ^{-1}\left(\frac{s-z_0}{\sqrt{X^2+Y^2}}\right)+8 s\bigg]\nn
-m^2 \left(\frac{16 z_0}{\left(X^2+Y^2\right)^2}-\frac{15}{8} \pi  \left(\frac{1}{X^2+Y^2}\right)^{3/2}\right)
\end{gather}
To calculate the other quantities we need to understand the perturbation of the null surface as well. The linear order calculation required the zeroth order expressions for the null generators, tangents vectors and embeddings. At quadratic order one also needs linear order corrections to these. In order to find these let us consider the linear order approximation to  the metric,
\begin{gather}
ds^2 = -\left(1+2\Phi\right) dt^2 
+\left( 1 -2\Phi\right)\left( dz^2 + dx^2+dy^2\right)
\end{gather}
with $\Phi=-\frac{ m }{ \sqrt{ \rho^2 + (z-z_0)^2 }}$. Let us also assume that the generators of the horizon gets perturbed to $l=\accentset{(0)}{l}+f$, which being null must satisfy,
%
\begin{gather}
2f^{\mathfrak a} \accentset{(0)}{l}_{\mathfrak a}+h_{\mathfrak a\mathfrak b}\accentset{(0)}{l}^{\mathfrak a} ~\accentset{(0)}{l}^{\mathfrak b}=\mathcal O(h^2)
\end{gather}
Further $l$ should be hypersurface orthogonal. A necessary condition for this to hold is that it must be geodesic. Since we are carrying out our calculation in the affine paramerisation we assume $l$ to be affinely parametrised, $\nabla_l~l=0+\mathcal O(h^2)$, which implies,
\begin{gather}
\accentset{(0)}{l}^{\mathfrak a}\accentset{0}{\nabla}_{\mathfrak a} f^{\mathfrak b}+C^{\mathfrak b}_{\mathfrak c\mathfrak d}\accentset{(0)}{l}^{\mathfrak c} ~\accentset{(0)}{l}^{\mathfrak d}+\accentset{(0)}{l}^{\mathfrak a}\accentset{0}{\nabla}_{\mathfrak a} \accentset{(0)}{l}^{\mathfrak b}=\mathcal O (h^2)
\end{gather}
Here, $C^{\mathfrak b}_{\mathfrak c\mathfrak d}$ indicates the perturbation of the Christoffel symbols. The null condition reduces to the following,
\begin{gather}
2(-f^t+f^z)-4\Phi=\mathcal O(h^2)
\end{gather}
while the geodesic condition gives,
\begin{gather}
\partial_tf^t+\partial_zf^t+2\partial_z\Phi=\mathcal O(h^2)\nn
\partial_tf^z+\partial_zf^z=\mathcal O(h^2)\nn
\partial_tf^x+\partial_zf^x+2\partial_x\Phi=\mathcal O(h^2)\nn
\partial_tf^y+\partial_zf^y+2\partial_y\Phi=\mathcal O(h^2)
\end{gather}
Since we have a metric for a freely falling particle in comoving coordinates, the only way that time dependence enters the picture is through the equation of the surface itself. The perturbations $f^{\mathfrak a}$ can therefore be taken to be independent of time. We thus have a consistent solution for the above sets of equations,
\begin{gather}
f^t=-2\Phi+B(x,y),~~f^z=B(x,y)\nn
f^x=-2\int\partial_x\Phi dz+F^x,~~~f^y=-2\int\partial_y\Phi dz+F^y,
\end{gather}
where $B(x,y)$ is some arbitrary function on the cross-sections. Let us now assume that there are two spacelike vectors $X_1=A_1\partial_t+B_1\partial_z+(1+C_1)\partial_x+D_1\partial_y$ and $X_2=A_2\partial_t+B_2\partial_z+(1+C_2)\partial_x+D_2\partial_y$ orthogonal to the null vector just found. They may not be orthonormal and will serve as coordinates for the cross sections. Note that we have chosen these vectors such that they are perturbations of $\partial_x$ and $\partial_y$.  The orthogonality of these with the null vector gives,
\begin{gather}
A_1=f^x+B_1+\mathcal O(m^2),~~A_2=f^y+B_2+\mathcal O(m^2)
\end{gather}
We need these vectors to commute with the null vector and with each other. This along with the geodesic condition is sufficient to ensure that $l$ is hypersurface orthogonal and that the three vectors span an integrable null submanifold.  These will serve as coordinate bases on the perturbed horizon. The commutation relations give the following sets of equations,
\begin{gather}
(\partial_zA_1-\partial_xf^t),~~\partial_zB_1=0,~~\partial_zC_1=\partial_xf^x, ~~\partial_zD_1=\partial_xf^y\\
(\partial_zA_2-\partial_yf^t),~~\partial_zB_2=0,~~\partial_zC_2=\partial_yf^x, ~~\partial_zD_2=\partial_yf^y\\
(\partial_xA_2-\partial_yA_1),~~\partial_xB_2-\partial_yB_1=0,~~(\partial_xC_2-\partial_yC_1)=0\nn
~~(\partial_yD_1-\partial_xD_2)=0
\end{gather}
To find the embedding functions we need to solve for the parameters defined as $\frac{\partial}{\partial s}=l$, $\frac{\partial}{\partial{X}}=X_1$, $\frac{\partial}{\partial{Y}}=X_2$
The solutions to the above equations, can be found in appendix (\ref{sol}). They have been obtained with specific boundary conditions, the choice of which determines the perturbed null vectors. It therefore directly influences the value of $\accentset{(1)}{\omega}^A$ calculated from these quantitites. However if $\accentset{(1)}{\omega}^A$ is calculated from the evolution equation then it is completely agnostic of this higher order calculation. Hence the boundary condition used in the evolution equation must be such that it matches with the expression calculated directly from these higher order quantities. This was the main motivation behind the particular choice of boundary condition made in the section (\ref{order1}). 

Since individual expressions will be big, we will explain,  what terms contribute at quadratic order, how they may be calculated, followed by the final expression for the change in the divergence. Let us consider the $\mathcal D^A\mathcal D^B\Sigma_{AB}$ term. This gets contibutions from terms of various orders in $\Sigma_{AB}$. Ordinary derivative acting on $\mathcal O(m^2)$ term of $\Sigma_{AB}$ and the covariant derivative (at $\mathcal O(m)$) acting on $\mathcal O(m)$ contribution to $\Sigma_{AB}$. Therefore to calculate this we need an expression for $\mathcal O(m^2)$ terms in $\Sigma_{AB}$. This is obtained by solving the evolution equation for $\Sigma_{AB}$ at quadratic order. This requires calculating quantities such as $g(l,C(l,X_1)X_1)$ (upto order $m^2$) followed by the use of the perturbed embedding functions, to restrict these expressions to the EH.

Another contribution will come from $\mathcal D^A(\omega^B\Sigma_{AB})$ term. Since both $\omega^A$ and $\Sigma_{AB}$ is of order $(m)$ the derivative here will be the usual flat space derivative. The term $\mathcal D^A\mathcal D_AK^{(l)}$ can be calculated by taking the flat space Laplacian on $K^{(l)}$. Using these, the evolution equation for the divergence may be solved.  Instead of calculating the instanteneous value we will find the total change in the divergence. This is convenient as it is independent of the boundary conditions used in the evaluation of the perturbed generators, tangent vectors and the embedding functions. The total change is found to be,
\begin{gather}
\Delta(div~\accentset{(2)}{\omega})\nn
=-\frac{8 m^2}{\big(X^2+Y^2\big)^2 \big(z_0^2+X^2+Y^2\big)^{7/2}} \Bigg[2 z_0^7+7 z_0^5 \big(X^2+Y^2\big)\nn
+8 z_0^3 \big(X^2+Y^2\big)^2
+6 z_0^2 \big(X^2+Y^2\big)^2 \sqrt{z_0^2+X^2+Y^2}\nn
+\big(X^2+Y^2\big)^3 \sqrt{z_0^2+X^2+Y^2}
+2 z_0^6 \sqrt{z_0^2+X^2+Y^2}\nn
+6 z_0^4 \big(X^2+Y^2\big) \sqrt{z_0^2+X^2+Y^2}+3 z_0 \big(X^2+Y^2\big)^3\Bigg]
\label{divomegelimit2}
\end{gather}
\begin{figure}[h]
	\includegraphics[width=8cm]{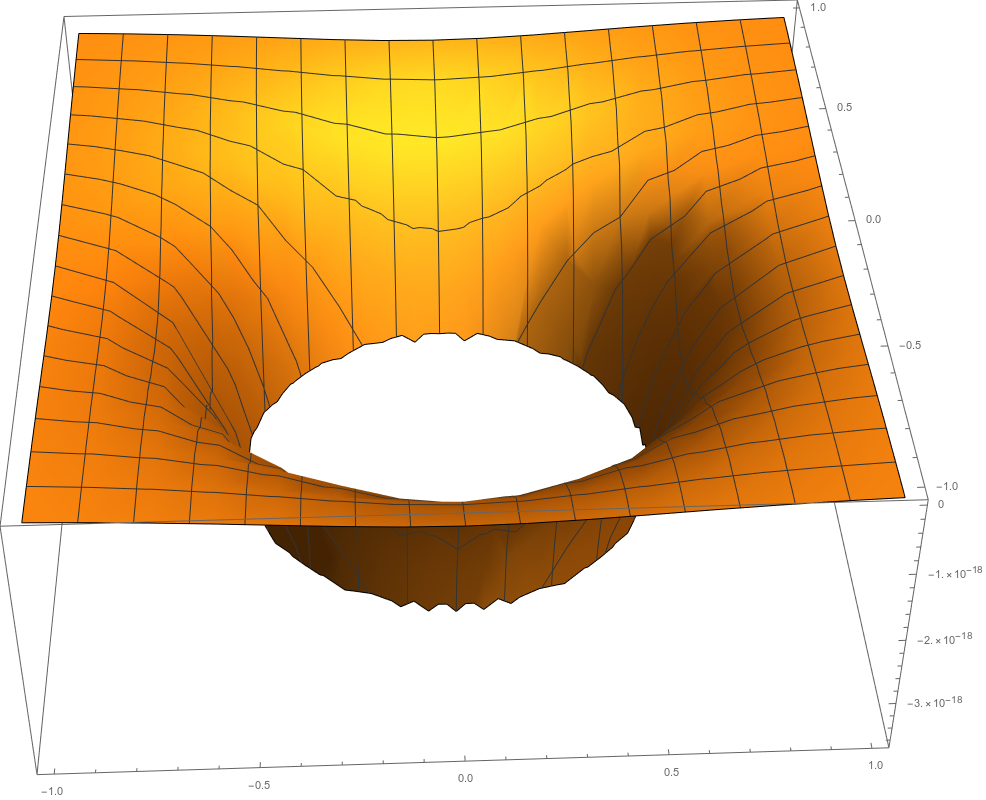}
	\includegraphics[width=8cm]{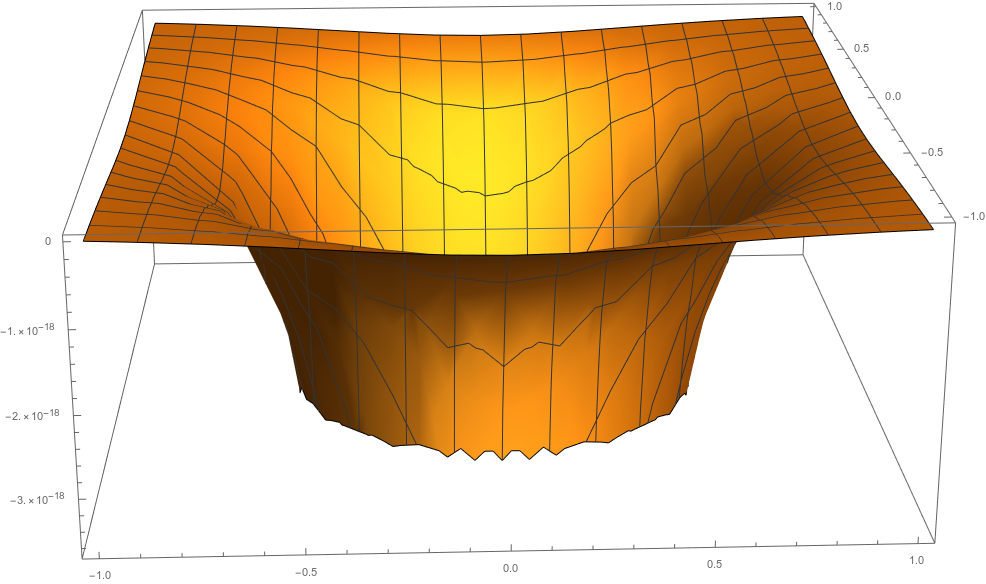}
	\caption{Plot of divergence of $\omega^A$ on the $x,y$ plane at thefinal slice $s\rightarrow 0$}
	\label{2divomega}
\end{figure}
A plot of the above expresion is shown in figure (\ref{2divomega}).
\section{Comment on further approximations}
At this point of time let us try to impose the restrictions on mass and $z_0$ that were assumed in \cite{Thorne:1986iy,Price:1986yy,Suen:1988kq}. Recall the expression for the tidal forces If we are interested in regions such that $X^2+Y^2<<z_0^2$, then the expressions can be replaced by delta functions,
\begin{gather}
g(\accentset{(0)}{l},\accentset{(1)}{C}(\accentset{(0)}{l},\partial_X)\partial_X) = - g(\accentset{(0)}{l},\accentset{(1)}{C}(\accentset{(0)}{l},\partial_Y)\partial_Y)\nn 
\approx-\frac{4 m (X-Y) (X+Y)}{\left(X^2+Y^2\right)^2}\delta(s-z_0)\nn
g(\accentset{(0)}{l},\accentset{(1)}{C}(\accentset{(0)}{l},\partial_Y)\partial_X)
\approx-\frac{8 m X Y}{\left(X^2+Y^2\right)^2}\delta(s-z_0)
\end{gather} 
Using these expressions the shear tensors can be calculated followed by the total change in the divergence at linear order. Interestingly the first order change in the divergence vanishes in this case. At quadratic order not all terms can be approximated by delta functions. Thus there is no point implementing it and a more complete solution is necessary as has been done in section (\ref{order2}). Taking the limit after performing the integrations e.g from eq. (\ref{divomegalimit}), shows that it is non zero. Hence it is clear that these two steps do not commute.  Taking the limit at the level of the source field is fine if one is interested in finding quantities that were of interest in \cite{Suen:1988kq,Amsel:2007mh,Bhattacharjee:2014eea}. However it is not a good idea to take this limit first where spatial derivatives w.r.t $X,~Y$ are involved, as in our case.
\section{Conclusion}
The motivation behind this work has been to check if supertranslations are indeed induced, for simplest of processes that are analytically tractable. The membrane para-digm provides us with a framework to deal with situations where the infalling object can be treated perturbatively. We have dealt with one such process that has been discuused before in the membrane paradigm framework, but in a different context. The results lead us to the conclsuion that supertranslations are indeed induced and motivates to check more involved processes. In particular, further investigations may tell us if there could be any observable signature like the memory effect in gravitational wave data, which is related to supertranslations at infinity. But this might require the utilisation of numerical tools.

It is quite widely accepted that as far as numerical relativity is concerned the \emph {Dynamical horizon} or locating marginally trapped surfaces is more suited for identifying Black hole horizons. But, as has been emphasized in the introduction, the Dynamical horizon framework is inherently non perturbative. Hence analytic calculation may only be possible in highly symmetric configurations where there is a significant simplification of the evolution equations. The drawback of such symmetry assumptions is that some terms are automatically zero due to the symmetry requirements \cite{Chatterjee:2020enf}.  The main aim of this paper was to study the effects of these terms. In particular the role of shear transpired in this calculation, whereas for spherically symmetric Dynamical horizons this term was trivial due to symmetry assumptions. Infact  in \cite{Chatterjee:2020enf} the only surviving term was a stress energy term. This indicated that supertranslations can be induced by the presence of a stress energy tensor which violates the dominant energy condition. Here the inclusion of the extra terms howver indicate that this may happen even in the absence of such pathological energy momentum tensor.

Let us now  comment on some of the subtleties in the current calculation. It should be noted that in general the evolution equation for the expansion does have a term which is linear in the mass of the perturbing body. This arises from the stress energy tensor of the body. But we have neglected the term and therefore the expansion is non zero only at quadratic order or higher.  Since the in-falling body is of finite extent there will always be some generators that do not pass through the interior of the body. Thus for these generators the stress energy term in the evolution equation for the expansion is indeed zero and also the tidal forces have the same expression as that in the exterior of a body. Our results hold only for these generators.
This is just a simplification as our main motivation was to see the effects of non-matter like terms.

To implement the above one must choose the parameters in the plots carefully.  Recall that a region of the plots was excised. This region if taken to be roughly of the same size as the body, automatically ensures the above condition to hold. $b$ which determines this size however has to be chosen appropriately. Note that due to the above mentioned artifact $div~\accentset{(2)}{\omega}$ diverges at $x=y=0$ and thus grows unboundedly around this point. $b$ therefore has to be so chosen in such a way that in remains a few orders of magnitude less than $div~\accentset{(1)}{\omega}$. 

Further, the span of coordinates $X$ and $Y$ should be such that the Rindler horizon remains valid. This range can be manipulated by appropriately choosing $M$. The freedom however is  not absoulute as $M$ satisfy a bound. In \cite{Suen:1988kq,Amsel:2007mh,Bhattacharjee:2014eea}, such a bound was obtained so that caustics do not form to the future of the bifurcation surface. In our notation this translates to an inequality between $b,~m$ and $M$. For the choices made here the bound turns out to be $M<\frac{3}{4\sqrt{2}}10^9$. Having pointed out these intricacies we must also mention that there is not much significance in exact numbers, obtained here. Rather it is the observation, that even in the simplest of processes supertranslations can be induced, that is important.

As an accompanying result we have moved beyond the linear order calculations which is the point of discussion in most of the membrane paradigm literature. A definite step wise algorithm has been introduced and followed to identify the event horizon beyond the zeroth order. This might be of importance for any future work on the membrane pardigm for black holes.

\section{Acknowledgements}
The author would like to thank Ayan Chatterjee for helpful discussions. The author acknowledges the support through grants from the NSF of China with Grant No: 11947301 and Fundamental Research Funds for Central universities under grant no.  WK2030000036.

\appendix
\section{Solving the PDEs for embedding functions}\label{sol}
In this section we will find the solutions fo the differential equation that we arrived at while trying to find the perturbed event horizon. Let us first try to find out the components of the perturbed null vector and the vectors $X_1,~X_2$ . The set of equations obtained from the geodesic condition and the commutation relation is sufficient for determining these vectors. The first of the commutation relations gives the following constraint on the function $B$.
\begin{gather}
(\partial_zA_1-\partial_xf^t)=-2\partial_z\int\partial_x\Phi dz+2\partial_x\Phi-\partial_xB=0
\end{gather}
which implies $\partial_xB=0$. Similarly,
$\partial_yB=0$. Thus we can take $B=0$. The expressions for the components of the vectors therefore reduce to,
\begin{gather}
f^t=-2\Phi,~~ f^z=0,~~f^x=-\int~ 2\partial_x\Phi ~dz+F^x(x,y)\nn
f^y=-\int~ 2\partial_y\Phi ~dz+F^y(x,y)
\end{gather}
The expression for the component $f^t$ is a results of the null condition while the expression for $f^x$ and $f^y$ has been obtained from the geodesic equation. Performing the integration leads to,
\begin{gather}
f^x= -\frac{2 m x (z-z_0)}{\left(x^2+y^2\right) \sqrt{(z-z_0)^2+x^2+y^2}}+F^x(x,y),
\end{gather}
where $F^x(x,y)$ is an integration constant.
The commutation relation involving $C_1$ and $f^x$ can now be intergrated to obtain $C_1$. Thus,
\begin{gather}
C_1=\int~\partial_xf^xdz+c_1(x,y)\nn
=-\frac{2 m \bigg((z-z_0)^2 \left(y^2-x^2\right)+x^2 y^2+y^4 \bigg)}{\left(x^2+y^2\right)^2 \sqrt{x^2+y^2+(z-z_0)^2}}+\partial_xF^x(x,y)z\nn
+c_1(x,y)
\end{gather}
But this expression  diverges as
$\bigg(\partial_xF^x(x,y)+\frac{2 m \left(x^2-y^2\right)}{\left(x^2+y^2\right)^2}\bigg)z$ in the limit $z\rightarrow\infty$. Thus we need to set,
\begin{gather}
F^x(x,y)=\frac{2 m x}{x^2+y^2}+\mathcal F^x(y)
\end{gather}
Further, an integration of the PDE involving $C_2$ and $f^x$ yields,
\begin{gather}
C_2=\int\partial_yf^xdz+c_2(x,y)\nn
=\frac{2 m x y \left(2 (z-z_0)^2-2 z\sqrt{(z-z_0)^2+x^2+y^2}+x^2+y^2\right)}{\left(x^2+y^2\right)^2 \sqrt{(z-z_0)^2+x^2+y^2}}\nn
+z  \partial_y\mathcal F^x(y)+c_2(x,y)
\end{gather}
This diverges as $z\partial_yF^x(y)$ as $z\rightarrow\infty$. Thus $F^x(y)$ has to be set to zero. By following similar arguments we have,
\begin{gather}
f^y=-\frac{2 m y (z-z_0)}{\left(x^2+y^2\right) \sqrt{(a-z)^2+x^2+y^2}}+\frac{2 m y}{x^2+y^2}
\end{gather}
\begin{gather}
D_2=\frac{2 m}{\left(x^2+y^2\right)^2} \Bigg((y^2-x^2) (\sqrt{x^2+y^2+(z-z_0)^2}-z)\nn
-\frac{y^2 \left(x^2+y^2\right)}{\sqrt{x^2+y^2+(z-z_0)^2}}\Bigg)+d_2(x,y)\nn
%
D_1=\frac{2 m x y}{\left(x^2+y^2\right)^2 \sqrt{(z-z_0)^2+x^2+y^2}} \Bigg(x^2+y^2\nn
+2(z-z_0)^2-2 z \sqrt{(z-z_0)^2+x^2+y^2} \Bigg)+d_1(x,y)
\end{gather}
The integration constants $c_1,~c_2,~d_1,~d_2$ must now be chosen such that they are consistent with boundary conditions.  Ideally in order  it is necessary to choose the integration constants in such a way that the perturbed bifurcation surface $s=0$ has the same area as the unperturbed black hole. Note that this kind of a boundary condition is necessary while writing a physical process first law. To this end let us try to find the metric induced on the horizon cross-section. 
\begin{gather}
ds^2=\bigg(1+\frac{2m}{\sqrt{X^2+Y^2+(s-z_0)^2}}+2C_1\bigg)(dX)^2\nn
\nn
+2(C_2+D_1)dX~dY\nn
+\bigg(1+\frac{2m}{\sqrt{X^2+Y^2+(s-z_0)^2}}+2D_2\bigg)dY^2
\end{gather}
We want this metric to have the same area two form as the flat unperturbed cross-section. A sufficient condition for this is to assume the following value for $C_1$ at the initial bifurcation surface,
\begin{gather}
\lim_{z\rightarrow\infty}C_1=-\frac{2 m \left(z_0^2 \left(y^2-x^2\right)+y^2 \left(x^2+y^2\right)\right)}{\left(x^2+y^2\right)^2 \sqrt{z_0^2+x^2+y^2}}+c_1(x,y)\nn
=-\frac{m}{\sqrt{z_0^2+x^2+y^2}}
\end{gather}
Similar result can be obtained for $D_2$. Invoking the commutation relation, involving $C_1,~C_2$ and $D_1,~D_2$, would then allow us to evaluate the integration constants for the remaining coeffecients. It can however be verified that the embedding functions obtained by using these boundary values are rather complicated making it difficult to obtain reasonable intermediate expressions during the calculation. A way out is to choose all these integration constants to be zero. This does not affect the value for the total change in the divergence of the rotation one form as it agnostic of such a choice. We therefore assume the simplest value for the these constants and evaluate the embedding functions. 
Recall the first of the PDEs determinining the embedding functions.
\begin{gather}
\frac{\partial t}{\partial s}=1-2\Phi=1+\frac{2m}{\sqrt{X^2+Y^2+(s-z_0)^2}}
\end{gather}
where we have used the zeroth iteration results in the right hand side. Thus
\begin{gather}
t=s+2 m \log \left(\sqrt{X^2+Y^2+(s-z_0)^2}+s-z_0\right)\nn
+g(X,Y)
\end{gather}
We choose $g(X,Y)=0$ for simplicity. The commutation relation are automatically satisfied. Integrating the equation for $x$ yields,
\begin{gather}
\frac{\partial x}{\partial s}=f^x\nn
x=\frac{2 m X \left(s-\sqrt{X^2+Y^2+(s-z_0)^2}\right)}{X^2+Y^2}+f(X,Y),
\end{gather}
where $f(X.Y)$ is the integration constant. Now using the expression for $\frac{\partial x}{\partial X}$,
\begin{gather}
\frac{\partial x}{\partial X}=1+C_1
\end{gather}
we get a constraint, $f=X+g(Y)$, on the integration constant. Invoking the equation for $\frac{\partial x}{\partial Y}$ implies $g(Y)=0$.
By following a similar procedure we have
\begin{gather}
y=\frac{2 m Y \left(s-\sqrt{X^2+Y^2+(s-z_0)^2}\right)}{X^2+Y^2}+Y
\end{gather}
\section{Notations and conventions}\label{N&C}
The covariant derivative $\nabla_WZ$ is a map $\nabla: T\mathcal M \otimes T\mathcal M\rightarrow T\mathcal M$, where $W,Z~\in ~T\mathcal M$. For any  is an immersed submanifold $\mathcal S$ in $\mathcal M$ the tangent space at a point $x\in\mathcal S$ can be decomposed into a space of vector tanget to $\mathcal S$ and a space of vector perpendicular to $\mathcal S$ viz. $T_x\mathcal M=T_x\mathcal S\oplus T_x^{\perp}\mathcal S$. The covariant derivative $\mathcal D_XY$ intrinsic to $\mathcal S$, where $X,Y~\in T\mathcal S$ can be obtained decomposing  the covariant derivative $\nabla$ on $\mathcal M$ as,
\begin{gather}
\nabla_XY=\mathcal D_XY+K(X,Y),
\end{gather}
where the extrinsic curvature $K(X,Y)$ in normal to $\mathcal S$ while $\mathcal D_XY$ is tangent to $\mathcal S$. In this paper the role of $\mathcal S$ will be played by the horizon cross-sections. If $X\in T\mathcal S$ and $N^\perp\in T^\perp\mathcal S $ then,
\begin{gather}
\nabla_XN^\perp=\nabla_X^\perp N^\perp-W_{N^\perp}(X).
\end{gather} 
Here  $\nabla^{\perp}_XN^\perp$ is the connection in the  normal bundle and is purely perpendicular to $\mathcal S$, while the shape operator $W_{N^\perp}(X)$ is tangent to $\mathcal S$. The extrinsic curvature and the shape operator are related through the following relation.
\begin{gather}
g(W_{N^\perp}(X),Y)=g(N^\perp,K(X,Y)),
\end{gather}
where $Y\in T\mathcal S$. The Riemann tensor on $\mathcal M$  and on $\mathcal S$ are defined as,
\begin{gather}
R(W,U)V\equiv[\nabla_W,\nabla_U]V-\nabla_{[W,U]}V\nn
\mathcal R(X,Y)Z\equiv[D_X,D_Y]Z-D_{[X,Y]}Z
\end{gather}
respectively.  Let $X,Y,Z,W\in T\mathcal S$ and $N^\perp\in T^\perp\mathcal S$. Then the Gauss and the Codazzi equations are given as,
\begin{gather}
g(R(X,Y)Z,W)=g(\mathcal R(X,Y)Z,W)\nn
-g(K(X,Z),K(Y,W))+g(K(X,W),K(Y,Z)),\\
g(R(X,Y)N^\perp,Z)=g((\nabla_YK)(X,Z),N^\perp)\nn
-g((\nabla_XK)(Y,Z),N^\perp)\label{Codazzi}
\end{gather}
respectively.

\end{document}